# Thermodynamics of G•A mispairs in DNA: continuum electrostatic model

**Julia Berashevich and Tapash Chakraborty***

*Department of Physics and Astronomy, University of Manitoba, Winnipeg, MB R3T 2N2, Canada. E-mail:tapash@physics.umanitoba.ca*

An analysis of the stability of a duplex containing G•A mispairs or G•A/A•G tandem during DNA melting has revealed that duplex stability depends on both DNA sequences and on the conformations of the G•A mispairs. The thermodynamics of single pair opening for G(anti)•A(syn) and G(anti)•A(anti) conformations adopted by G•A mispairs is found to strongly correlate with that of the canonical base pairs, while for sheared conformation a significant difference is observed.

## 1 Introduction

Various external agents and the cell metabolism are known to be responsible for DNA damage which is directly related to many human diseases including cancer. The mispairs are the most frequent DNA damage that are not always effectively recognized and repaired by the enzymatic repair system [1]. Dependence of the recognition mechanism and its recognition efficiency on the type of mispair remains unclear as yet, but recently it was justified that repair enzymes activate the base pair opening and flipping of a nucleoside out of the DNA helix [2-4]. The G•A mismatch is one of the most poorly recognized mismatches [1,5] in DNA. The insignificant distortion induced by this mismatch into the DNA duplex cast doubt on the initial proposal of the enzymatic recognition of the skeleton distortion. On the other hand, it has generated a lot of interest on the experimental study of thermodynamic stability of the DNA duplex containing a single G•A mismatch (denoted as G in the following) [6] and a more structurally stable G•A/A•G tandem (denoted as GA in the following) [7-11]. The only conclusion that can be drawn from all the earlier studies is that the DNA sequence influences the G•A mismatch stability. However, the results obtained by different groups have shown rather contradictory behavior that is yet to be explained. In particular, the same changes in the DNA sequences were shown to cause a decrease of DNA stability in some cases and an increase in another [7-11]. The molecular dynamics simulations of the DNA duplex containing sheared G•A mismatch have shown that the duplex creates a stable conformation to incorporate this mismatch [12].

It is known that depending on the surrounding environment, such as the DNA sequences and the solvent p**H**, the G•A mismatch can adopt different conformations: G(anti)•A(anti), G(anti)•A(syn), G•A sheared and G(syn)•A(anti) [5,7,10]. Thus, the molecular dynamics simulations have demonstrated that depending on the sequence context the G•A/A•G tandem can adopt different conformations [13]. Therefore, we suspect that the reason for the contradictory observation in the experiments [7-11] is in fact, due to different mispair conformations. To clear up this issue we performed theoretical simulations of DNA melting and computed the thermodynamic stability of the duplexes containing G•A mismatches in different conformations. Because the experimental data are available largely for the G•A/A•G tandem as the most stable conformation, we devoted our work to a thorough investigation of the tandem adopted G(anti)•A(anti), G(anti)•A(syn) and sheared conformations, whose thermodynamic stability has been extensively investigated experimentally. However, the single G•A mismatch is known to be the most frequent damage and we analyze its behavior within the DNA duplex as well. For the tandem, we obtained an opposite influence of the DNA sequence on the thermodynamics of duplex and tandem formation if the tandem adopts the G(anti)•A(anti), G(anti)•A(syn) conformations as opposed to the G•A sheared conformation. Moreover, the calculated thermodynamics of the single mispair opening within a DNA duplex, which is however difficult to measure experimentally, surprisingly indicates that the G(anti)•A(anti) and G(anti)•A(syn) conformations are characterized by the same thermodynamics as that for a canonical A-T pair and depends weakly on the nearest sequences. We also compared our results with those for G•T and A•C [14] mismatches that are efficiently recognized by the repair proteins [1] and found its thermodynamics to be significantly different from that of any canonical pairs. Based on these results we believe that the thermodynamics of pair opening inside the DNA duplex can be the first and most important step for proofreading and recognition processes. The recently discovered flipping of the nucleosides out of the DNA helix during the repair process [2-4] is clearly an additional evidence in support of the important role of thermodynamics of pair dissociation in the recognition process.

## 2 Computational method

The reaction of the base pair formation is

$$base + base \xrightarrow{\Delta G_s} pair, \qquad (1)$$

where $\Delta G_s$ is the standard free energy of reaction in the solvent. According to the continuum electrostatic model [15], $\Delta G_s$ can be calculated within the thermodynamic cycle through the standard free energy of reaction in the gas phase $\Delta G_g^T$ and the energy shift required to transfer this reaction from the gas phase to the molecule environment $\Delta E^{g \to s}$, i.e.

$$\Delta G_s = \Delta G_g^T + \Delta E^{g \to s} \qquad (2)$$

The same procedure can be applied to determine the free energy of the pair formation in the aqueous solution. The energy shift to transfer the reaction (Eq. 1) from the gas phase to the solution is $\Delta E^{g \to a}$.

We begin our investigation of the G•A mismatch geometries in vacuum and later make a comparison of the thermodynamics of their pair formation with that of the canonical base pairs in vacuum and inside the DNA duplex. The standard free energy for pair formation in the gas phase at temperature $T$ can be expressed as

$$\Delta G_g^T = E_p^T - E_b^T \qquad (3)$$

where $E_p^T$ and $E_b^T$ are the energies of the base pair and the sum of the energies of the separated bases, respectively.

For the calculation of the standard free energy for pair formation in the gas phase (see Eq. 3) the geometries of the bases and base pairs were optimized in vacuum with the quantum-chemistry methods within the Jaguar 6.5 program [16]. These computations were based on the density-functional theory using the Becke3-Lee-Yang-Parr functional [17]. The restricted basis set with polarization and diffuse functions 6-31++G** has been applied. The optimized geometries were used to compute $E_p^T$ for the base pairs and $E_b^T$ for the separated bases. The values of these energies are determined as a sum of the zero-point electronic energy and the vibrational components. The enthalpy $\Delta H_g^T$ of the base association is estimated as the difference between the whole pair enthalpy and those for the separated bases, and is similar to the expression for the free energy in Eq. 3. The basis set superposition error has been included into the calculation of $\Delta G_g^T$ and $\Delta H_g^T$ by the counterpoise method using the individual bases as fragments [18].

It has been shown earlier [12, 14, 19] that the density functional theory (DFT) and Hartree-Fock (HF) methods have significant discrepancy of the hydrogen bond length for optimized geometries of the DNA base pairs that subsequently modifies the calculated data for the thermodynamics of the base pair formation. We compared the hydrogen bond lengths and their energies with two different methods B3LYP/6-31++G**//B3LYP/6-31++G** functional and HF/6-31++G**//HF/6-31++G**. According to our results, the length of the hydrogen bonds calculated within HF for the G•A mispair and for the canonical pairs is longer up to ~ 0.11-0.20 Å than those with the DFT methods (see values of the hydrogen bonds for the G•A mispairs in Figure 2.). These results are in excellent agreement with the results of Refs. 14 and 19. Therefore, the thermodynamics of the base pair formation is also different for these two methods. Thus, $\Delta G_g^{298K}$ calculated with HF/6-31++G**//HF/6-31++G** is -7.38 kcal/mol for the A-T pair, -20.81 kcal/mol for the G-C pair, -8.35 kcal/mol for the G(anti)•A(syn), -9.09 kcal/mol for the G(anti)•A(anti) and 4.58 kcal/mol for the G•A sheared. These values are shifted by ~2.5 kcal/mol to that obtained with the B3LYP/6-31++G**//B3LYP/6-31++G**. However, the relation between these data for the different base pairs within one method is the same for both DFT and HF. However, our DFT results are found to agree very well with the experimental data and with computed data in [20] performed with MP2/6-31G*(0.25)//HF/6-31G** and in Refs. 14,19 performed with BP86/TZ2P//BP86/TZ2P. Therefore, taking into account all the points presented above, we can conclude that the B3LYP/6-31++G**//B3LYP/6-31++G** functional used in our work properly describes the thermodynamic properties of base pair in the vacuum and their geometries, that was also concluded for the BP86/TZ2P method in Ref. 14.

In the solvent, the DNA molecule and therefore the base context, strongly associates with the solvent due to several types of interactions. First of all, the DNA interacts with the water molecules [21] and with the sugar-phosphate backbone carrying the negative charges, the positively charged ions from the solvent are attracted and accumulated along the DNA phosphate. Secondly, the bases interact with partial charges of the nearest bases [22]. Therefore, in comparison to the vacuum, the solvent produces a shift of the free energy of base association reaction due to these interactions and due to the changes in the dielectric environment. Therefore, the free energy of base association in the aqueous solution $\Delta G_a$ and within the solute DNA molecule $\Delta G_s$ becomes:

$$\Delta G_{a(s)} = \Delta G_g + (\Delta E_p^{g \to a(s)} - \Delta E_b^{g \to a(s)}). \qquad (4)$$

The energy shift arising due to the transfer of association reaction from the vacuum into the aqueous solvent $\Delta E^{g \to a}$ or into the molecular environment $\Delta E^{g \to s}$ can be taken into account within the continuum electrostatic model [23,24]. The restrained electrostatic potential procedure (RESP) [25] has been applied to calculate the atomic partial charges for the electrostatic calculations of the bases and base pairs after their geometry optimization within the quantum chemical methods in vacuum. For the calculation of $\Delta E^{g \to a}$ the optimized geometries (in vacuum) of the separated nucleobases and nucleobase pairs with the individual atomic partial charges obtained within the RESP have been placed into the homogeneous continua – the vacuum and the aqueous solution. For the calculation of $\Delta E^{g \to s}$ of the G•A mispair and the GA tandem within the DNA duplex, the DNA sequences have been generated with the CHARMM program [26] using the previously optimized (with quantum chemical methods) structure of the G•A mispair adopted different conformations. To build the DNA sequences using the CHARMM program the structural parameters were obtained from *1bna* [27]. The DNA duplexes containing the single G•A mispair and GA tandem were generated with help of structural parameters for the sugar-phosphate backbones obtained in works [27-29]. The obtained DNA duplexes, where base pairs geometries were previously optimized with quantum chemical method, have been also placed into the two different environments – the vacuum and the aqueous solution during the electrostatic computations.

The APBS program with a three dimensional Poisson-Boltzmann solver was employed [30] for the calculations of the $\Delta E^{g \rightarrow a}$ and the $\Delta E^{g \rightarrow s}$ energies. In computations of the DNA duplexes and single DNA strand obtained due to the melting process, the electrostatic intrastrand interactions between stacked base pairs and interstrand interactions between opposite strands were taken into account. Because the DNA molecule is highly charged in the solvent environment, the linear solution of the Poisson-Boltzmann equation is not applicable, and a nonlinear form has been used. Within the electrostatic model, the DNA molecule was represented as a continuum with a low dielectric constant, individual atomic partial charges and the van der Waals radii [31]. The solvent has been represented as a homogeneous continuum with a high dielectric constant. The following dielectric constants have been used: for vacuum and for the molecular groups, $\varepsilon = 1$, while for the solvent, $\varepsilon = 78.3$. The high resolution grid with the step of 0.25 Å centered at the evaluated base pair or nucleobases has been applied. The solvent radius was 1.4 Å, the ionic strength was 0.1 M and the temperature was 298 K. The cubic B-spline charge discretization method and multiple Debye-Hückel boundary conditions have been applied.

## 3 Pair formation in vacuum and in aqueous solution

We begin our investigation of the G•A mismatch geometries in vacuum and later make a comparison of the thermodynamics of their pair formation with that of the canonical base pairs in vacuum and inside the DNA duplex. We computed the $\Delta G_g^T$ (free energy) and $\Delta H_g^T$ (enthalpy) at room temperature for association of adenine and thymine, guanine and cytosine, adenine and guanine in the vacuum with quantum-chemical methods. The G(anti)•A(anti), G(anti)•A(syn) and G•A sheared conformations adopted by the G•A mispairs have been considered. Our results for the ionization potential (IP) determined within the Koopman's theorem and the thermodynamics for pair formation of the canonical pairs and for the G•A mismatches are presented in Table I. We found that the guanine inside a G•A mispair is characterized by a much higher IP than inside the G-C pair, but is still lower than that of adenine inside the A-T pair. For the base association, the enthalpy and the free energy of the canonical pair formations differ for the A-T and G-C pairs by ~ 10kcal/mol. For the G•A mismatch, the thermodynamics of formation of the G(anti)•A(syn) and G(anti)•A(anti) conformations is in fact, similar to that for the canonical A-T pair. A large difference from the A-T pair is observed only for the G•A sheared conformation. A comparison of our results for $\Delta H_g^{298K}$ with the experimental data for the canonical pairs shows very good agreement.

Table I: Properties of the canonical pairs and G•A mismatches in vacuum at temperature $T=298$ K computed with B3LYP/6-31++G**//B3LYP/6-31++G**. All values are in kcal/mol

| pair | IP | $\Delta G_g^{298K}$ | $\Delta H_g^{298K}$ | $\Delta H^{298K\ a}$ |
|---|---|---|---|---|
| A-T | 141.96 | -10.16 | -10.75 | -12.1 |
| G-C | 125.88 | -22.46 | -23.05 | -21.0 |
| G(anti)•A(syn) | 139.96 | -11.31 | -11.90 | - |
| G(anti)•A(anti) | 137.47 | -11.82 | -12.41 | - |
| G•A sheared | 137.67 | -7.57 | -8.16 | - |

[a] Experimental data [32] with correction proposed in [33].

The consideration of the base association reaction within the energy cycle takes care of the interactions of the bases with the surrounding environment listed above. The first step is to estimate the base association in the aqueous solution. The $\Delta E_b^{g \rightarrow a}$ and the $\Delta E_p^{g \rightarrow a}$ energies have been computed as electrostatic solvation energies and their calculated values are presented in Table II. The aqueous solution shifts equally the free energies of base association for the canonical base pairs, compared to the energies in vacuum, and the energy difference for these pairs is ~9 kcal/mol. The association energy $\Delta G_a$ of adenine and thymine in the aqueous solution is not strongly negative, that makes the A-T pair less stable thermodynamically than G-C. The magnitudes of $\Delta G_a$ have shown excellent agreement with the experimental data that indicates the suitability of our model for the following simulations of the base association in the solute DNA. We found that the G•A mispair is certainly unstable within the aqueous solution. However, the melting temperature of DNA duplex (even containing the mismatches) is much higher than the room temperature [6-11], which makes DNA stable in our cells. Therefore, the estimation of the base association energy within the DNA molecule is our main concern for this work. For these purposes we generated the DNA helices containing the mismatches (for details of helix generation see sec. Computational method). Since we found some interesting structural features of the DNA helix due to incorporation of the mismatch, this issue is discussed prior to the base pair energetics within the DNA molecule.

Table II. Association of the bases in the aqueous solution at temperature $T=298$ K. All values are in kcal/mol.

| pair | $\Delta E_b^{g \rightarrow a}$ | $\Delta E_p^{g \rightarrow a}$ | $\Delta G_a$ | $\Delta G_a^{\ a}$ |
|---|---|---|---|---|
| A-T | -24.26 | -15.35 | -1.25 | -1.15 [34] |
| G-C | -37.23 | -25.19 | -10.42 | -11.0 [35] |
| G(anti)•A(syn) | -31.45 | -20.43 | -0.29 | - |
| G(anti)•A(anti) | -31.45 | -18.52 | 1.11 | - |
| G•A sheared | -31.45 | -23.99 | -0.11 | - |

[a] Experimental data

## 4 Structural changes of DNA skeleton induced by G•A/A•G tandem

The incorporation of the mispairs can cause a local or a global structural instability of the helix (Ref. [36] and references therein) and can change its thermodynamical stability. For the G•A mismatch, the single G•A mispair was earlier found to destabilize the DNA structure [7, 28], while the adjacent G•A/A•G mismatches stabilize the DNA helix [7, 37]. We noticed that the structural destabilization of DNA induced by the single G•A mispairs can be attributed to their spatial geometry. Since within the G•A pair the nucleobases are located in the planes with different slopes related to each other, that was also found in Ref. 12, the G•A geometry is non-planar, unlike the almost flat geometries of the canonical pairs. The term "flat geometry" is used when tilt of the bases within the pairs is close to 0 (for example for the canonical base pairs within B-DNA the tilt is -1°). The G•A sheared mispair is found to have the most non-planar geometry.

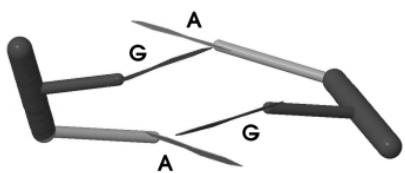

**Fig. 1.** The side view of the 5´-(GA)-3´ tandem in the G•A sheared conformation created with the CHARMM [26]. The single G•A mispair has a non-planar geometry which destabilizes the DNA duplex. The tandem structure, on the other hand, compensates the influence of the nonplanarity to the DNA skeleton by placing two mispairs in reverse order. The top view of this structure is available in Fig. 2 (c).

The destabilization of the DNA skeleton by incorporation of the G•A mispair can, in fact, be prevented by placing two G•A mispairs in reverse order, which would compensate for the non-planarity of a single mispair (Fig. 1). Therefore, the GA tandem induces mostly local perturbation of the phosphate backbone and exhibits better structural stability. The conformation G(anti)•A(syn) adopted by the GA tandem is found to induce minor distortion to the B-DNA helix, because the planar size of the G•A pair is similar to the canonical pairs. In Fig. 2, three different conformations of the G•A mispair adopted by the GA tandem structure are presented. We have measured the distance between the two carbon atoms (C1´…C1´) attaching bases to the sugar-phosphate backbone on the opposite strands. For the canonical A-T pair consisting of three aromatic rings, the C1´…C1´ distance is 10.48 Å. For the G•A mispair containing four rings the C1´…C1´ distance is 11.046 Å for the G(anti)•A(syn) (Fig. 2(a)), 13.084 Å for the G(anti)•A(anti) (Fig. 2 (b)) and 8.806 Å for the G•A sheared (Fig. 2 (c)). The G(anti)•A(syn) conformation, whose planar size is close to the canonical pair size, should induce less distortion in the DNA helix. Because the G(anti)•A(anti) has a large planar size, incorporation of this conformation will cause significant stretching of the phosphate backbone, whose elasticity is limited, connecting the mispairs and the neighboring base pairs. In contrast, the G•A sheared conformation actually compresses the phosphate backbone due to its smaller planar size. Therefore, the G(anti)•A(syn) and G•A sheared conformations seems to be more capable of forming a stable DNA duplex. However, the location of the pairs within the GA tandem in sheared conformation significantly differs from that for other conformations and also for the canonical pairs (see Fig. 2 (c)). The twist angle between the G•A and A•G mispairs within the sheared conformation adopted by a tandem is close to 90° against the ~25° for the G(anti)•A(syn) and G(anti)•A(anti) conformations and against the 36° for the regular B-DNA structure. Therefore, for the G•A sheared conformation the guanines (or the adenines) belonging to the G•A and to the A•G mispairs are located at the top of each other (see Fig. 2 (c)), while for other conformations the guanines are located above or below the adenines (see Fig. 2 (a-b)). To adopt the tandems within the DNA duplexes, the twist angle between a G•A mispair and the nearest-neighbor canonical pair must be ~ −15° for the sheared conformation and ~ 45° for the other two. Therefore, for the sheared conformation the strong interstrand interaction occurs between nucleobases of the same type (guanine/guanine), while for the G(anti)•A(syn) and G(anti)•A(anti) conformations, in opposite, between nucleobases of different type (guanine/adenine). Because of these structural peculiarities, the 5´-C(GA)G-3´ duplex with G•A in sheared conformation is characterized by a lower energy than the 5´-G(GA)C-3´, and vice versa for the G(anti)•A(anti) and G(anti)•A(syn) conformations. These results have been obtained by comparing the total energies of the structures (without geometry optimization) performed within the quantum chemical methods in vacuum.

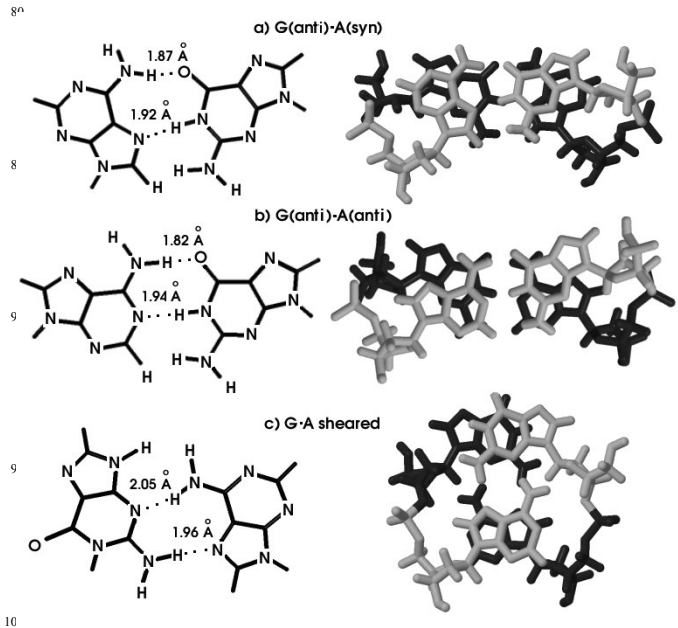

**Fig.2.** Structures of the 5´-(GA)-3´ sequences created with the CHARMM [26] where the tandem adopts different conformations. The two pairs are distinguished by their different shades. The scale is the same for all figures to show the noticeable difference in the planar size of the conformations. For the numerical values of the planar sizes and the twist angles, see the text. The geometries of the G•A mispairs optimized with the quantum chemistry method (B3LYP/6-31++G**//B3LYP/6-31++G**) are presented on the left hand side. The obtained length of the hydrogen bonds between bases is shown for the all conformations.

## 5 Stability of G•A/A•G tandem in molecular environment

Since the GA tandem is structurally stable, the problem of the thermodynamic stability of the DNA duplex containing a tandem has attracted a lot of attention [6-11]. We have already mentioned that the observed thermodynamics of the duplex formation has been very contradictory, and in different experiments the same change in DNA sequence has led to an increase [10-11] or a decrease [7-9] of the duplex stability. An explanation of this phenomenon has not been found yet and the next part of our work will be devoted to the solution of this puzzle. Since the experimental data are available mostly for the GA tandem, in the following we consider that structure for detail investigations and comparison with the experimental data, while the single G•A mispair incorporated into the DNA molecule will be briefly analyzed towards the end.

The next step is to estimate the energetic of the G•A mispairs within the solute DNA. Due to the transfer of the G•A mispairs from vacuum into the solution, the IP is shifted on a $\Delta E_{pair}^{g \to s}$. Our computational results for the energy $\Delta E_{pair}^{g \to s}$ of different conformations adopted by G•A are presented in Table III. According to these results, the G•A sheared conformation demonstrates the lowest $\Delta E_{pair}^{g \to s}$ energy because of the tandem geometry where the strong electrostatic interactions between the nucleobases occur (see Fig. 2 (c)). Among all the sequences, the purine – GA – pyrimidine and the purine – (TA) – pyrimidine combinations exhibit the lowest value of $\Delta E_{pair}^{g \to s}$. For a single G•A mispair incorporated into the DNA duplex, the lowest value of this energy corresponds to the purine – G – purine sequences. Therefore, stacking of guanines from the G•A pair with other purines in the same strand reduces the ionization potential of the G•A mispair. Obviously, incorporation of the G•A mispairs into the DNA structure changes the energetic conditions for hole migration due to the difference of the G•A ionization potential from that of the canonical pairs. Therefore, degradation of the charge transfer rate can provide information about the presence of mismatch in the DNA structure and even their conformations. This procedure was proposed earlier for the design of the bio-sensors capable of detecting the mutated or damaged DNA [38].

The thermodynamic stability of the GA tandem in solution has been simulated as melting of the DNA duplex $\Delta G_{duplex}^{m}$, which provides us with the opportunities to make a direct comparison of our computational results with the experimental data [7-11]. We have computed the change of the electrostatic interaction between the nucleobases due to the separation of the DNA duplex $\Delta E_{p}^{g \to s}$ into two single strands $\Delta E_{b}^{g \to s}$. The base conformations are the same for the double-stranded and the single-strand structures. It is assumed that the negative charge on the phosphate backbone is neutralized by the solvent ions. Because we found that only the nearest-neighbor base pairs to that for GA tandem significantly contributes to $\Delta E_{pair}^{g \to s}$ while the effect of the others base pairs has been estimated to be only ±0.03 eV, for the calculations of the thermodynamics stability of the DNA duplex only nearest-neighbor canonical base pairs have been taken into account.

The free energy of the DNA duplex formation ($\Delta G_{duplex}^{m}$) containing the GA tandem or the (TA) pairs incorporated between the G-C pairs are presented in Table III. For structures containing the (TA) pairs and the GA tandem in the G(anti)•A(syn) and G(anti)•A(anti) conformations, the more thermodynamically stable duplexes correspond to pyrimidine – GA – purine sequences, as has been observed experimentally [7-9]. For the duplexes with the GA tandem in sheared conformation, we have found an opposite behavior, where the purine – GA – pyrimidine sequence shows the lowest value of $\Delta G_{duplex}^{m}$, that is in good agreement with the other experimental results [11]. It is known that the type of pair conformation within the GA tandem depends significantly on the pH of the solution (Ref. [10] and references therein) and the nearest sequences [7]. Therefore, based on our computational analysis, we can explain the observed contradiction in the first [7-9] and the second [11] cases by relating it to different G•A mispair conformations.

Table III. The thermodynamics of the duplex ($\Delta G_{duplex}^{m}$), tandem ($\Delta G_{tandem}^{m}$) and the G•A mispair ($\Delta G_{pair}^{m}$) formation simulated from DNA melting and the energy of a single G•A formation inside a DNA duplex ($\Delta G_{pair}^{h}$). All values are in kcal/mol.

| 5′- … -3′ | $\Delta E_{pair}^{g \to s}$ [a] | $\Delta G_{duplex}^{m}$ | $\Delta G_{tandem}^{m}$ | $\Delta G_{pair}^{m}$ [a] | $\Delta G_{pair}^{h}$ [a] |
|---|---|---|---|---|---|
| Canonical T-A pairs ($\Delta G_g$ =-10.16 kcal/mol) | | | | | |
| C(TA)G | -9.58 | -32.99 | -13.85 | -6.90 | -9.64 |
| C(TA)C | -9.04 | -28.13 | -12.25 | -6.48 | -9.68 |
| G(TA)C | -8.77 | -21.28 | -9.98 | -4.99 | -9.89 |
| G(TA)G | -9.31 | -27.22 | -12.01 | -5.77 | -9.86 |
| G(anti)•A(syn) conformation ($\Delta G_g$ =-11.31 kcal/mol) | | | | | |
| C(GA)G | -13.14 | -20.28 | -6.75 | -3.60 | -10.24 |
| C(GA)C | -14.09 | -16.70 | -5.85 | -4.94 | -10.18 |
| G(GA)C | -14.96 | -9.48 | -1.48 | -0.95 | -10.02 |
| G(GA)G | -13.90 | -16.58 | -5.70 | -1.41 | -10.10 |
| G(anti)•A(anti) conformation ($\Delta G_g$ =-11.82 kcal/mol) | | | | | |
| C(GA)G | -8.85 | -26.76 | -12.23 | -5.83 | -10.36 |
| C(GA)C | -9.54 | -23.21 | -8.44 | -7.53 | -10.30 |
| G(GA)C | -12.60 | -17.28 | -6.00 | -3.20 | -10.30 |
| G(GA)G | -11.90 | -20.72 | -8.63 | -1.31 | -10.44 |
| G•A sheared conformation ($\Delta G_g$ =-7.57 kcal/mol) | | | | | |
| C(GA)G | -18.58 | -11.26 | -3.66 | -2.07 | -6.81 |
| C(GA)C | -19.22 | -17.76 | -8.56 | -3.22 | -6.96 |
| G(GA)C | -21.22 | -30.33 | -13.59 | -7.02 | -7.37 |
| G(GA)G | -20.58 | -21.94 | -8.64 | -5.82 | -7.22 |

[a] The $\Delta E_{pair}^{g \to s}$, $\Delta G_{pair}^{m}$ and $\Delta G_{pair}^{h}$ energies correspond to the first G•A pair from the GA tandem

In a DNA duplex, the entire geometry of the duplex contributes to the $\Delta G_{duplex}^{m}$ energy. However, more important is the part of this energy that describes the formation of the GA tandem or G•A mispair within this tandem, which is difficult to measure experimentally. The corresponding $\Delta G_{tandem}^{m}$ and $\Delta G_{pair}^{m}$ energies are presented in Table III. We have found that the influence of the DNA context on the formation energy of the tandem $\Delta G_{tandem}^{m}$ has the same impact as that for duplex formation $\Delta G_{duplex}^{m}$. The dipole moments of each G•A mispair arising from the partial charge distribution are oriented in opposite direction within the tandem structure that produces an energetic compensation. Because of this compensation, the maximum and minimum values of $\Delta G_{pair}^{m}$ for the G•A mispair within the tandem can be shifted with respect to $\Delta G_{tandem}^{m}$. We also analyzed the $\Delta G_{pair}^{m}$ for a duplex containing not the tandem, but a single G•A mispair. As expected, for the G(anti)•A(syn) and G(anti)•A(anti) conformations the lowest $\Delta G_{pair}^{m}$ corresponds to the pyrimidine – G – pyrimidine sequence and for the G•A sheared conformation, to the purine – G – purine sequence, in agreement with the experiment [10]. Clearly, determination of the mispair conformation should be the prime concern in the experimental study of the mismatch stability.

It should however be noted that the DNA melting is a process of strand separation, while for proofreading and repair processes, a single pair dissociation within the duplex is the key procedure [2, 3]. It is known that the hydrogen bonds are channels for charge exchange between the nucleobases during the pair formation [14]. Therefore, we have calculated the energy $\Delta G_{\text{pair}}^m$ required to process the charge exchange between separate nucleobases to form a pair within the DNA sequences. We have obtained a weak dependence of $\Delta G_{\text{pair}}^m$ for one G•A mispair from the tandem on the sequence context (see results in Table III). Moreover, the solvent decreases the energy difference for dissociation of the stacked (TA) pairs and the <u>GA</u> in G(anti)•A(syn) and G(anti)•A(anti) conformations in comparison to that in vacuum, and the discrepancy of the magnitude of $\Delta G_{\text{pair}}^m$ for these pairs is minor. For the sheared conformation this discrepancy is larger for both vacuum and solvent. Therefore, the incorporation of the G•A mismatch into the DNA duplex do not significantly change the energy required to open this mismatch from that in vacuum, and hence, the energy of the pair formation in vacuum $\Delta G_g$ can be used as a crude estimation for the thermodynamics of pair formation in vacuum and within a DNA duplex. To prove this conclusion we performed calculations for the G•T mispair, as presented in the Table IV. We choose the G•T mispair conformation characterized by $\Delta G_g^{298K}$ similar to that for the A-T pair (see GT2 in Refs. 14,20). Clearly, the free energy of the base association reaction for the G•T mispair, which results are presented in Table IV, is different from that for the A-T pairs by ~ 2.5 kcal/mol.

Table IV. The $IP$ of the G•T mispair and the thermodynamics of the pair formation in the vacuum $\Delta G_g^{298K}$ (B3LYP/6-31++G**//B3LYP/6-31++G**), aqueous solution $\Delta G_a$ and inside the 5´- G(G•T)C -3´ duplex $\Delta G_{\text{pair}}^h$. All values are in kcal/mol.

| Planar G•T geometry | Energy | Value |
|---|---|---|
| 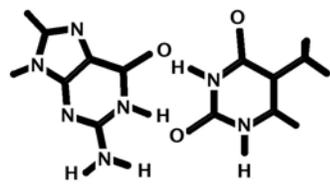 | $\Delta G_g^{298K}$ | -12.85 |
| | $IP$ | 134.88 |
| | $\Delta G_a$ | -2.10 |
| | $\Delta G_{\text{pair}}^h$ | -12.39 |

## 6 Discussion and Conclusion

Originally, several hypotheses for mismatch recognition were proposed for proofreading and repair processes. These include recognition of the backbone structural changes, the nucleobase context and local DNA dynamics [28]. Recent discovery of flipping of the bases out of the DNA helix activated by the repair proteins UDG (Uracil DNA glycosylase) [2-3] has motivated new view of this problem. At present, the recognition process can be divided into several steps: pair dissociation, base flipping out of the helix, subsequent interaction of the flipped base with the enzymes, and base association back to the pair. The repair enzymes are found to be able to catalyze the base pair opening [3]. As the energy required to open the A-T pair is ~ 10 kcal/mol and for the G-C pair ~ 20 kcal/mol, the enzymes must apply two different catalytic conditions for pair-opening in these two cases. The G•A, T•C, T•T mispairs are known to be less efficiently recognized by the enzymes (Ref. 1 and reference therein), and moreover they have the same pair dissociation thermodynamics as that for the canonical base pairs [14,20]. However, the G•T and A•C mispairs, whose formation energies in vacuum lie in the middle of $\Delta G_g^{298K}$ of the A-T and G-C pairs (see the Table IV for the G•T), are known to be easily recognized. Based on these thermodynamical data, we expect that the enzymatic recognition starts from the pair dissociation and detection of pairs that have thermodynamic properties distinctively different from that of the canonical pairs. For example, for G•T and A•C mispairs (~ 14 – 15 kcal/mol [14,20]), the enzymes need to apply the pair opening energy of G-C, and the leftover energy goes to change the base-flipping dynamics. It is known that the ATP hydrolysis is used to unzip a DNA molecule [39], and perhaps can also be used for pair opening in mispair recognition, as follows: To open the A-T pair, a single hydrolysis is necessary, but the G-C pair opening requires more than one hydrolysis reactions [40]. Therefore, the processes of opening and closing of the canonical base pairs preserve the initial concentration of the reactant and products of the hydrolysis reaction, while in the case of G•T and A•C mispairs the number of ATP agents will be reduced after closing of these pairs.

For the mispairs that escape recognition during the pair opening, the enzymes can apply the subsequent recognition of the flipped nucleoside inside an active-site pocket [4]. Therefore, for pairs that dissociate like the A-T pair (viz., the G•A and T•C mispairs), if the flipped nucleobase differs from adenine or thymine then the pair will be recognized as a mispair. Hence, the T•C mispair should be recognized highly efficiently, while recognition of the G•A pair depends on the type of the flipped nucleoside. For the pairs that dissociate like the G-C pair the nucleobase other than guanine or cytosine will be recognized as belonging to a mispair.

To summarize, the proposed recognition mechanism is derived from the finding that for the G•A mismatch adopted particular conformations, the thermodynamics of pair opening (see the - $\Delta G_{\text{pair}}^h$ energy in Table III) is similar to that for the canonical A-T pair and therefore, the thermodynamics is the reason for poor recognition of this mismatch by the repair enzymes activating the pair dissociation. The large discrepancy of the thermodynamics of pair opening for the sufficiently recognized mismatches such as G•T (see - $\Delta G_{\text{pair}}^h$ in the Table IV and - $\Delta G_g^{298K}$ in Ref. 14 and 20) and A•C (see - $\Delta G_g^{298K}$ in Ref. 14 and 20) is additional evidence to support our proposal.

## 7 Acknowledgement

We wish to thank Dr. Tom Brown (Southampton) for valuable suggestions in preparing the manuscript. The work was supported by the Canada Research Chairs Program and NSERC Discovery Grant.